# New type of half-metallic fully compensated ferrimagnet

S. Semboshi,[1] R.Y. Umetsu,[1,2,3,*] Y. Kawahito,[4,5] H. Akai[6,*]

[1]Institute for Materials Research, Tohoku University; 2-1-1 Katahira, Sendai 980-8577, Japan.
[2]Center for Spintronics Research Network, Tohoku University; 2-1-1 Katahira, Sendai 980-8577, Japan.
[3]Center for Science and Innovation in Spintronics; 2-1-1 Katahira, Sendai 980-8577, Japan.
[4]Advanced Science-Technology Research Program, Japan Agency for Marine-Earth Science and Technology; 2-15 Natsushima, Yokosuka, Kanagawa 237-0061, Japan.
[5]Department of Applied Physics, Graduate School of Engineering, Osaka University; 2-1 Yamadaoka, Suita, Osaka 565-0871, Japan.
[6]The Institute for Solid State Physics, The University of Tokyo; 5-1-5 Kashiwanoha, Kashiwa, Chiba 277-8581, Japan.

*Corresponding authors. Email: rieume@imr.tohoku.ac.jp (R.Y. Umetsu), akai@issp.u-tokyo.ac.jp (H. Akai).

**Abstract**
Half-metallic fully compensated ferrimagnets (HM-FCFMs), a special class of half-metals exhibiting zero magnetization at absolute zero, are promising candidates for next-generation spintronics applications. For over 25 years, theoretical studies have been conducted to realize HM-FCFM materials for practical applications. Herein, we experimentally demonstrate a NiAs-type hexagonal-structured (CrFe)S compound that could serve as an HM-FCFM material. It has a half-metallic nature, with 100% spin-polarized Fermi surfaces and zero magnetization. Further, the magnetization shows a linear behavior as a function of the magnetic field below the compensation temperature of around 200 K, with high magnetic coercivity of 38 kOe at 300 K. These magnetic features are expected to contribute to a quantum leap in the application of HM-FCFM layers in spintronics devices.

**Main Text**

Spintronic data storage devices, such as giant magneto resistors (GMR) and tunneling magneto resistors (TMR), are essential tools for quantum information processing and probabilistic computing [1]. In spintronic materials, the spins of electrons must be controlled in addition to electronic charges. In particular, a 100% spin-polarized current source is ideal for application in advanced spintronic devices [2]. Materials with 100% spin polarization can be realized using half-metals [3], which behave as metals in one spin direction and insulators or semiconductors in the other. However, such half-metals need to be designed through atomic-level engineering, and only a few half-metals have been developed thus far. Some examples include Heusler alloys [4-6]; transition metal oxides such as rutile [7], spinel [8], and perovskite compounds [9]; and diluted magnetic semiconductors [10, 11].

All half-metals that have been reported thus far are ferromagnetic. However, half-metals with other magnetic characteristics could lead to the development of new spintronic devices. One such case, proposed by van Leuken and de Groot in 1995 [12] is half-metallic fully compensated ferrimagnets (HM-FCFM). These are commonly called half-metallic antiferromagnets; however, antiferromagnets inherently cannot be half-metallic because their spin rotational symmetry is not compatible with half-metallicity. In contrast, while fully compensated ferrimagnets, in which the electron spins of crystallographically inequivalent magnetic sublattices align antiparallel to each other with zero net magnetic moment, behave magnetically like antiferromagnets [13, 14], they can be used as half-metals. A significant feature of HM-FCFMs is that no static magnetic field is generated within them; hence, there is low magnetostatic energy as the total magnetic moment is diminished at temperatures near 0 K. Thus, HM-FCFMs would be useful for generating a 100% spin-polarized current in a zero-field environment, facilitating the development of novel spintronic devices that differ from those prepared through conventional methodologies. To this end, theoretical calculations to identify HM-FCFM materials have been intensively conducted for over 25 years, with a number of possible HM-FCFM materials and their properties being predicted [15–27]. However, most of them are non-practical or appear only in unstable crystal structures. Only $Mn_{1.5}V_{0.5}FeAl$ bulk Heusler alloys [28, 29] and $Mn_2V_{0.5}Co_{0.5}Al$ melt-spun ribbons [30] with $L2_1$-type crystal structures have demonstrated $N$-type ferrimagnetism in fully compensated ferrimagnets, as predicted by Néel [13, 14]. The band structure calculations for $Mn_{1.5}V_{0.5}FeAl$ suggest that it has a half-metallic electronic structure [29].

Akai and colleagues [27, 31] predicted the possible existence of pnictide and chalcogenide HM-FCFM materials with NiAs-type structures on the basis of first-principles electronic structure calculations performed using the Korringa–Kohn–Rostoker (KKR) method. They concluded that for transition metal magnetic systems to act as HM-FCFMs, the total number of $d$-electrons per magnetic ion must equal 5 (i.e., the number of $d$-electrons for a pair of magnetic ions must be 10). Accordingly, some pnictides and chalcogenides with hexagonal NiAs-type structures such as (CrFe)S, (CoV)Sb, and Cr/Fe-doped ZnS are predicted to be HM-FCFM materials although none of them have been synthesized yet. Among the predicted candidates, (CrFe)S is attractive as it composed of common elements with high Clarke numbers. In addition, the Cr–Fe–S phase diagram indicates that it could be obtained via a conventional metallurgical process (**Fig. S1**). If (CrFe)S is evidenced to exhibit the beneficial HM-FCFM behavior, it would have a significant impact on its practical application toward the development of spintronics devices. Therefore, in this study, we attempted to synthesize NiAs-type chalcogenide (CrFe)S and evaluate its magnetic properties and half-metallicity.

While (CrFe)S has a relatively wide compositional range, as shown in the Cr–Fe–S phase diagram (**Fig. S1**), (CrFe)S with an Fe/Cr ratio of 1.0 satisfies the $d$-electron number requirement. To obtain single-phase (CrFe)S with HM-FCFM properties, we synthesized a slightly S-rich compound with an Fe/Cr ratio of 1.0 (($Cr_{23}Fe_{23})S_{54}$ at.%) by sintering raw elemental powders (see Supplementary Methods). The synthesized compound did not appear to contain a second phase or inclusions and was composed of typical crystal grains in which the constituent elements were mixed homogeneously (**Fig. 1A**). All peaks in the X-ray diffraction pattern of the sample (**Fig. S2A**) were indexed to a hexagonal NiAs-type (CrFe)S lattice (space group $P6_3/mmc$ (No. 194)), with the Cr and Fe atoms occupying the Wyckoff $2a$ positions (0, 0, 0) and the S atoms occupying the $2c$ sites (1/3, 2/3, 1/4), as illustrated in **Fig. 1B**.

It is difficult to elucidate details of the actual crystal structure of the off-stoichiometric (CrFe)S sample, such as the atomic ordering and vacancy substitution, through X-ray diffraction (XRD) analysis. However, Sokolovich and Bayukov [32] reported that a stoichiometric (CrFe)S compound synthesized by sintering and quenching exhibited a NiAs-type disordered structure, rather than a CdI$_2$-type ordered structure (**Fig. 1C**), on the basis of observations from Mössbauer spectra. Furthermore, when comparing the experimental XRD profile (**Fig. S2A**) with simulated profiles for NiAs-type (CrFe)S structures (**Figs. S2B-E**), the experimental peak intensities for off-stoichiometric (CrFe)S agree well with the simulated intensities for the structure containing vacancies at the 2*a* sites together with Cr and Fe atoms.

Moreover, differential scanning calorimetry measurement (DSC; **Fig. S3**) indicated that the NiAs-type (CrFe)S phase is stable below approximately 700 K, as predicted by the phase diagram (**Fig. S1**). **Figure S3** also indicates that the Curie temperature ($T_C$) of the sample is ~500 K, as there is a small step increment at this temperature during the first heating process.

Next, we evaluated the magnetic properties of the synthesized (CrFe)S sample. The thermomagnetization (*M–T*) curves are shown in **Fig. 2A**. Under a low magnetic field (200 and 500 Oe), magnetization decreases with an increase in temperature, until it crosses the zero value at 200 K (i.e., compensation temperature, $T_{comp}$). This is followed by a convex downward shape in the higher temperature region, with convex upward behavior just below $T_C$. This behavior is typical of *N*-type ferrimagnets [13, 14], except for crossing the zero value and the convex upward behavior just below $T_C$. $T_C$ was measured to be around 500 K, as indicated by the arrow in **Fig. 2A**, which corresponds to the value measured by DSC (**Fig. S3**). Under a high magnetic field (5 and 10 kOe), magnetization does not cross the zero value, and the upward convex behavior observed below $T_C$ gradually becomes prominent. This feature reflects the behavior of *P*-type ferrimagnets. The temperature dependence of magnetization is attributed to the presence of two magnetic sublattices. Because the magnetizations of the two sublattices compete in these ferrimagnets, the *M–T* curves vary depending on the strength of the applied magnetic field [13, 14].

The magnetization (*M–H*) curves at temperatures ranging from 6 to 425 K are shown in **Fig. 2B**. Relatively low magnetization (<4 emu/g) is observed in all *M–H* curves even under a high magnetic field of 90 kOe. Further, the shape of the *M–H* curves is temperature-dependent. At low temperature (6–100 K (**Fig. 2B(a)**), the curves exhibit a nearly linear behavior with small hysteresis. At 6 K, the *M–H* curve deflects at around ±70 kOe, suggesting a slight change in the magnetic ordering state. However, for measurements under 50 kOe, the *M–H* curves are linear without hysteresis. Over 150–200 K (below $T_{comp}$ of 200 K) (**Fig. 2B(b)**), the *M–H* curves appear as perfectly straight lines, characteristic of a material with fully compensated ferrimagnetic properties [13, 14]. In contrast, at 250–325 K (above $T_{comp}$) (**Fig. 2B(c)**), a hard ferromagnetic behavior is observed through the *M–H* curves. This temperature range corresponds to the convex downward trend in the *M–T* curves. The magnetic coercivity ($H_{cr}$) reaches a maximum at 300 K, with a value of 38 kOe, which decreases as the temperature increases further (**Fig. 2B(d)**).

The experimentally determined relationship between temperature and $H_{cr}$ is summarized in **Fig. 2C**. Although the value of $H_{cr}$ is ~19 kOe at 250 K, this may be an underestimation because the *M–H* curve does not saturate, even at 90 kOe. The energy product, *BH*, is relatively small, as magnetization is low because of the ferrimagnetic properties of the material. The high $H_{cr}$ value of 38 kOe at 300 K for the (CrFe)S sample, which does not contain rare-earth or noble-metal elements, is noteworthy. This value is competitive with those exhibited by $L1_0$-Mn$_{1.5}$Ga epitaxial films at room temperature [33] and ε-Rh$_x$Fe$_{2-x}$O$_3$ nanoparticles at 200 K (42.8 and 42 kOe, respectively) [34]. The high magnetocrystalline anisotropies of $L1_0$-MnGa and ε-Fe$_2$O$_3$ were revealed by theoretical calculations [35, 36]. Although the origin of the high magnetic anisotropy of the as-prepared compound is unclear, $H_{cr}$ could be increased with optimization. Such a large magnetic coercivity above $T_{comp}$ is also very interesting and facilitates application for spintronic device development.

The synthesized (CrFe)S samples must exhibit not only fully compensated ferrimagnetic properties but also half-metallicity. To determine the existence of half-metallicity, we investigated the total and partial density of states (DOS) of the stoichiometric CdI$_2$-type ordered and NiAs-type disordered (CrFe)S compounds by first-principles calculations (**Figs. 3A** and **3B**, respectively) (see Supplementary Methods). Here, the partial DOS implies the existence of an angular-momentum-decomposed DOS within each Voronoi cell surrounding a constituent atom. If the (CrFe)S sample has stoichiometric ordered or disordered structures, the Fermi level ($E_f$) is located in the deep band gap for the spin-down state, while it is located within the band for the spin-up state. This indicates that the (CrFe)S system has a half-metal-type electronic state. Further, the total magnetic moment of the system is almost zero, which proves the fully compensated ferrimagnetism of the system. In this study, we have confirmed that the half-metal-type electronic state is observed when the Fe/Cr ratio in the (CrFe)S compound is 1.0 even if the lattice parameters fluctuate slightly and the S content deviates from the stoichiometric composition (**Figs. S4** and **S5**). This suggests that the S-rich off-stoichiometric (CrFe)S sample synthesized in this study is half-metallic and the measured magnetic moment is adequately small to exhibit compensated ferrimagnetism. Therefore, we conclude that the half-metallicity of the (CrFe)S system is robust against these types of compositional and structural variations. Because the (CrFe)S system is half-metallic, the properties of fully compensated ferrimagnetism are also preserved. The deviation of spin polarization from 100% given in **Fig. 3** is due to a non-zero imaginary part attached to the Fermi energy, which is necessary for the numerical calculations.

The physical properties of the ordered CdI$_2$-type and disordered NiAs-type (CrFe)S systems, which were calculated using the optimized or experimental lattice constants, are listed in **Table 1**. In the NiAs-type disordered (CrFe)S system with optimized lattice constants, the spin magnetic moments of Cr and Fe are 3.27 and −3.06 $\mu_B$, respectively. As mentioned in the case of partial DOS, the electronic states of both Cr and Fe atoms are localized, resulting in large magnetic moments. The magnitudes of the magnetic moments are comparable and may allow the existence of fully compensated ferromagnetism even in a disordered state. The total energy of the CdI$_2$-type ordered structure is ~300 meV lower than that of the NiAs-type disordered structure, suggesting that the former state is more stable in the ground state. In the present experiments, however, the sample was obtained by quenching from 1253 K, in order to avoid phase decomposition and to obtain a single-phase (CrFe)S compound (see Supplementary

Methods). Therefore, obtaining a disordered state should not be surprising because the difference in the total energies between the ordered and disordered states is relatively small. The Curie temperature, $T_C$, was calculated by mean field approximation using the exchange interaction, $J$, which in turn was calculated by Liechtenstein's method [37]. It is generally known that $T_C$ calculated by mean field approximation is typically larger than the corresponding experimentally determined value if the thermal contribution is not considered. Under identical conditions, the $T_C$ of the NiAs-type disordered compound was ~200 K lower than that of the CdI$_2$-type ordered compound. The spin polarization of the CdI$_2$-type ordered compound was calculated to be 99.7%, indicating the existence of complete half-metal-type electronic states. Furthermore, high spin polarization is retained even in the disordered state and with some fluctuation of the lattice parameters, implying that the atomic distribution of Fe and Cr has no effect on half-metallicity. This would be a significant advantage for various applications.

In theory, NiAs-type (CrFe)S acts as an electron conductor with one spin orientation (up-spin, shown in red in **Fig. 4A**), but as an insulator with the opposite spin orientation (down-spin, shown in blue in **Fig. 4A**), which leads to the generation of current with high polarity. This feature has attracted interest as it facilitates potential applications in spintronics. For example, compared to conventional TMRs, which contain ferromagnetic (e.g., permalloy) inner and outer layers (**Fig. 4B**), TMR devices containing an HM-FCFM layer are expected to have drastically improved TMR ratios, as the leakage current would be minimized. The Julliére model suggests that the TMR ratio in magnetic tunneling junctions is infinity when complete half-metal ferromagnets are used as the electrode layers [38]. This not only results in drastic improvement of performance but also leads to the construction of the TMR device with a simple structure. Additionally, it is noteworthy that (CrFe)S is composed of common elements with high Clarke numbers. Therefore, this innovation is expected to provide a quantum leap in the application of HM-FCFM materials in spintronic devices.

In summary, despite attempts for over 25 years toward the development of HM-FCFM materials, only the Mn$_{1.5}$V$_{0.5}$FeAl bulk Heusler alloy has been reported as a potential HM-FCFM candidate thus far [29]. Here, we successfully synthesized a NiAs-type (CrFe)S compound and confirmed its behavior as a new type of HM-FCFM material. The HM-FCFM (CrFe)S material was developed on the basis of the $d$-electron number rule proposed by Akai and colleagues [27, 31]. Note that the Mn$_{1.5}$V$_{0.5}$FeAl bulk Heusler alloy also satisfies the $d$-electron number requirement. Thus, the $d$-electron number rule should be a principal guideline for fabricating new HM-FCFM families in the future. We are confident that the findings of this study will trigger new innovation in the spintronics field.

**Acknowledgments:** We thank Dr. N. Yodoshi and Mr. I. Narita of IMR, Tohoku University for their help with the experiments and Ms. Mori of JAMSTEC for assistance with the illustrations. We also acknowledge the support from the Cooperative Research and Development Center for Advanced Materials (CRDAM) of IMR, Tohoku University.
**Funding:** JSPS KAKENHI grant (18H01743).
**Author contributions:** Conceptualization: SS, RU, YK, HA; Investigation: SS, RU, HA; Writing – original draft: SS; Writing – review & editing: RU, YK, HA.
**Competing interests:** Authors declare that they have no competing interests.
**Data and materials availability:** All data are available in the main text or the supplementary materials.

**References**

1. W. A. Borders, A. Z. Pervaiz, S. Fukami, K. Y. Camsari, H. Ohno, S. Datt, Integer factorization using stochastic magnetic tunnel junctions. *Nature* **573**, 390-393 (2019).
2. S. A. Wolf, D. D. Awschalman, R. A. Buhrman, J.M. Daughton, S. von Molnár, M.L. Roukes, A. Y. Chtchelkanova, D.M. Treger, Spintronics: A spin-based electronics vision for the future. *Science* **294**, 1488-1495 (2001).
3. R. A. de Groot, F. M. Mueller, P. G. van Engen, K. H. J. Buschow, New class of materials: Half-metallic ferromagnets. *Phys. Rev. Lett.* **50**, 2024-2027 (1983).
4. J. Kübler, First-principle theory of metallic magnetism. *Physica B+C* **127**, 257-263 (1984).
5. J. Kübler, A. R. Williams, C. B. Sommers, Formation and coupling of magnetic moments in Heusler alloys. *Phys. Rev. B* **28**, 1745-1755 (1983).
6. S. Ishida, S. Akazawa, Y. Kubo, and J. Ishida, Band theory of $Co_2MnSn$, $Co_2TiSn$, and $Co_2TiAl$. *J. Phys. F* **12**, 1111-1122 (1982).
7. J. M. D. Coey, M. Venkatesan, Half-metallic ferromagnetism: Example of $CrO_2$. *J. Appl. Phys.* **91**, 8345-8350 (2002).
8. A. Yamase, K. Shiratori, Band structure in the high temperature phase of $Fe_3O_4$. *J. Phys. Soc. Jpn.* **53**, 312-317 (1984).
9. K.-I. Kobayashi, T. Kimura, H. Sawada, K. Terakura, Y. Tokura, Room temperature magnetoresistance in an oxide material with an ordered double-perovskite structure. *Nature* **395**, 677-680 (1998).
10. H. Akai, Ferromagnetism and its stability in the diluted magnetic semiconductor (In, Mn)As. *Phys. Rev. Lett.* **81**, 3002-3005 (1998).
11. M. Shirai, Electronic and magnetic properties of $3d$ transition-metal-doped GaAs. *Physica E* **10**, 143-147 (2001).
12. H. van Leuken, R. A. de Groot, Half-metallic antiferromagnets. *Phys. Rev. Lett.* **74**, 1171-1173 (1995).
13. L. Néel, Magnetic properties of ferrites: Ferrimagnetism and antiferromagnetism. *Ann. Phys.* **12**, 137-198 (1948).
14. L. Néel, Magnetism and local molecular field. *Science* **174**, 985-992 (1971).
15. W. E. Pickett, Spin-density-functional-based search for half metallic antiferromagnets. *Phys. Rev. B* **57**, 10613-10619 (1998).
16. V. Pardo, W. E. Pickett, Compensated magnetism by design in double perovskite oxides. *Phys. Rev. B* **80**, 054415 (2009).
17. H. Wu, Electronic structure study of double perovskites $A_2$FeReO$_6$ ($A$ = Ba, Sr, Ca) and Sr$_2$$M$MoO$_6$ ($M$ = Cr, Mn, Fe, Co) by LSDA and LSDA+U. *Phys. Rev. B* **64**, 125126 (2001).
18. Min Sik Park, B. I. Min, Electronic structures and magnetic properties of La$A$VMoO$_6$ ($A$ = Ca, Sr, Ba): Investigation of possible half-metallic antiferromagnets. *Phys. Rev. B* **71**, 052405 (2005).
19. Y. K. Wang, G. Y. Guo, Robust half-metallic antiferromagnets La$A$Mo$Y$O$_6$ ($A$ = Ca, Sr, Ba; $Y$ = Re, Tc) from first-principles calculations. *Phys. Rev. B* **73**, 064424 (2006).
20. S. T. Li, Z. Ren, X. H. Zhang, C. M. Cao, Electronic structure and magnetism of $Mn_2CuAl$: A first-principles study. *Physica B* **404**, 1965-1968 (2009).



21. I. Galanakis, K. Özdoğan, E. Şaşıoğlu, B. Aktaş, *Ab initio* design of half-metallic fully compensated ferrimagnets: The case of $Cr_2MnZ$ ($Z$ = P, As, Sb, and Bi). *Phys. Rev. B* **75**, 172405 (2007).
22. I. Galanakis, K. Özdoğan, E. Şaşıoğlu, and B. Aktas, Ferrimagnetism and antiferro-magnetism in half-metallic Heusler alloys. *Phys. Stat. Sol. (a)* **205**, 1036-1039 (2008).
23. H. Luo, Z. Zhu, G. Liu, S. Xu, G. Wu, H. Liu, J. Qu, Y. Li, Ab-initio investigation of electronic properties and magnetism of half-Heusler alloys $X$CrAl ($X$ = Fe, Co, Ni) and NiCr$Z$ ($Z$ = Al, Ga, In). *Physica B* **403**, 200-206 (2008).
24. E. Şaşıoğlu, Nonzero macroscopic magnetization in half-metallic antiferromagnets at finite temperatures. *Phys. Rev. B* **79**, 100406R (2009).
25. H. Akai, M. Ogura, Half-metallic diluted antiferromagnetic semiconductors. *Phys. Rev. Lett.* **97**, 026401 (2006).
26. M. Nakao, Tetrahedrally coordinated half-metallic antiferromagnets. *Phys. Rev. B* **74**, 172404 (2006).
27. M. Ogura, C. Takahashi, H. Akai, Calculated electronic structures and Neel temperatures of half-metallic diluted anti-ferromagnetic semiconductors. *J. Phys.: Condens. Matter* **19**, 365226 (2007).
28. R. Stinshoff, A. K. Nayak, G. H. Fecher, B. Balke, Completely compensated ferrimagnetism and sublattice spin crossing in the half-metallic Heusler compound $Mn_{1.5}FeV_{0.5}Al$. *Phys. Rev. B* **95**, 060410(R) (2017).
29. R. Stinshoff, G. H. Fecher, S. Chadov, A. K. Nayak, B. Balke, S. Ouardi, T. Nakamura, C. Felser, Half-metallic compensated ferrimagnetism with a tunable compensation point over a wide temperature range in the Mn-Fe-V-Al Heusler system. *AIP Adv.* **7**, 105009 (2017).
30. P. V. Midhunlal, J. A. Chelvane, D. Prabhu, R. Gopalan, N. H. Kumar, $Mn_2V_{0.5}Co_{0.5}Z$ ($Z$ = Ga, Al) Heusler alloys: High $T$c compensated P-type ferrimagnetism in arc melted bulk and N-type ferrimagnetism in melt-spun ribbons. *J. Magn. Magn. Mater.* **489**, 165298 (2019).
31. N. H. Long, M. Ogura, H. Akai, New type of half-metallic antiferromagnet: Transition metal pnictides. *J. Phys.: Condens. Matter* **21**, 064241 (2009).
32. V. V. Sokolovich, O. A. Bayukov, Mossbauer spectra of $Fe_xCr_{1-x}S$ solid solutions. *Phys. Solid State* **49**, 1920-1922 (2007).
33. L. Zhu, S. Nie, K. Meng, D. Pan, J. Zhao, H. Zheng, Multifunctional $L1_0$-$Mn_{1.5}Ga$ films with ultra-high coercivity, giant perpendicular magnetocrystalline anisotropy and large magnetic energy product. *Adv. Mater.* **24**, 4547-4551 (2012).
34. S. Ohkoshi, K. Imoto, A. Namai, S. Anan, M. Yoshikiyo, H. Tokoro, Large coercive field of 45 kOe in a magnetic film based on metalsubstituted ε-iron oxide. *J. Am. Chem. Soc.* **139**, 13268-13271 (2017).
35. A. Sakuma, Electronic structures and magnetism of CuAu-type MnNi and MnGa. *J. Magn. Magn. Mater.* **187**, 105-112 (1998).
36. M. Yoshikiyo, K. Yamada, A. Namai, S. Ohkoshi, Study of the electronic structure and magnetic properties of ε-$Fe_2O_3$ by first-principles calculation and molecular orbital calculations. *J. Phys. Chem. C* **116**, 8688-8691 (2012).
37. A. I. Liechtenstein, M. I. Katsnelson, V. P. Antropov, V. A. Gubanov, Local spin density functional approach to the theory of exchange interactions in ferromagnetic metals and alloys. *J. Magn. Magn. Mater.* **67**, 65-74 (1987).



38. M. Julliére, Tunneling between ferromagnetic films. *Phys. Lett.* **54A**, 225-226 (1975).
39. ASM Alloy phase diagram database, https://matdata.asminternational.org/apd/index.aspx.
40. M. Ogura, H. Akai, The full potential Korringa–Kohn–Rostoker method and its application in electric field gradient calculations. *J. Phys.: Condens. Matter*. **17**, 5741-5756 (2006).
41. H. Akai, Electronic structure Ni-Pd alloys calculated by the self-consistent KKR-CPA method. *J. Phys. Soc*. Jpn. **51**, 468-474 (1982).
42. H. Akai, Fast Korringa-Kohn-Rostoker coherent potential approximation and its application to FCC Ni-Fe systems. *J. Phys: Condens*. Matter **1**, 8045-8064 (1989).
43. J. P. Perdew, K. Burke, M. Ernzerhof, Generalized gradient approximation made simple. *Phys. Rev. Lett.* **77**, 3865-3868 (1996).


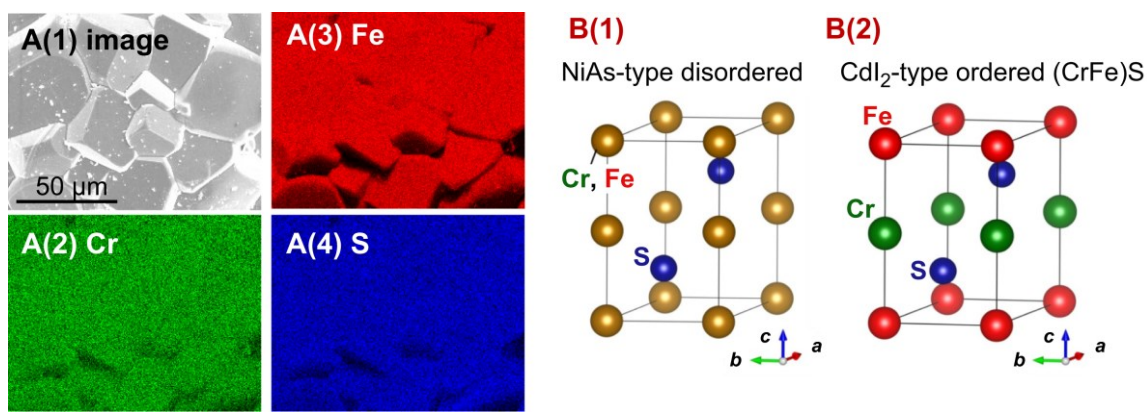

**Fig. 1. Structural characterization.** (**A**) Scanning electron microscopy–energy dispersive X-ray spectroscopy of the synthesized (CrFe)S sample: (**A(a)**) Backscattered electron image and elemental mapping for (**A(b)**) Cr, (**A(c)**) Fe, and (**A(d)**) S, taken from the same field of view as (A(a)). The synthesized compound comprised typical crystal grains with a size of ~40 μm. The constituent Cr, Fe, and S elements seem to be homogeneously mixed; the dark contrast in (A(b–d)) is caused by surface irregularities. (**B**) NiAs-type disordered structure (space group $P6_3/mmc$ (No. 194)) and (**C**) CdI$_2$-type ordered structure (space group $P3m1$ (No. 156)). The crystal structure of the off-stoichiometric (CrFe)S sample matched the NiAs-type disordered structure, with excessive S atoms, with vacancies (*Va*) and Cr and Fe atoms randomly occupying the 2*a* sites (see Supplementary Methods). The lattice parameters were measured to be $a = 0.3445$ and $c = 0.5744$ nm.

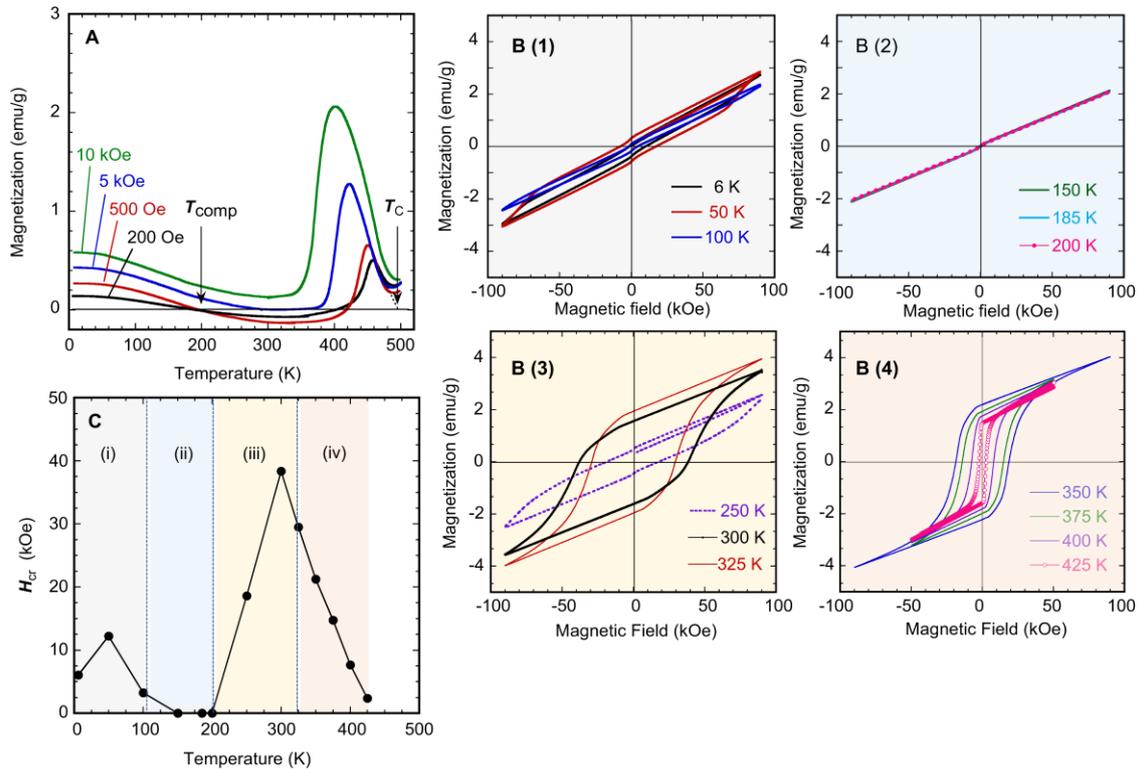

**Fig. 2. Magnetic properties.** (**A**) Thermomagnetization ($M$–$T$) curves of the (CrFe)S sample measured at 200 Oe, 500 Oe, 5 kOe, and 10 kOe. (**B**) Magnetization ($M$–$H$) curves at different temperatures, which show different behavior. (**B(a)**) 6–100 K: Small hysteresis in the full-loop; (**B(b)**) 150–200 K: Perfectly linearly relationship, indicating a fully compensated ferrimagnetic state; (**B(c)**) 250–325 K (above $T_{comp}$ of 200 K): Abrupt hard magnetization behavior; and (**B(d)**) 350–425 K: Magnetic coercivity ($H_{cr}$) gradually decreases with increasing temperature. (**C**) Relationship between temperature and magnetic coercivity ($H_{cr}$) at magnetic fields up to 90 kOe. Regions (i–iv) correspond to B(a–d).

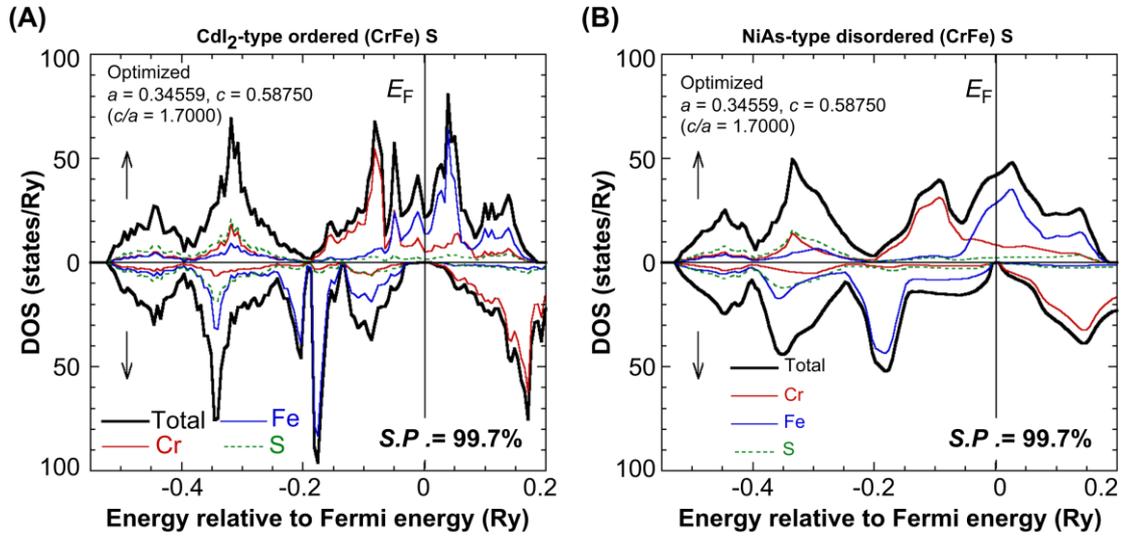

**Fig. 3 Density of states (DOS).** Partial DOS of Cr, Fe, and S and the total DOS of (**A**) CdI$_2$-type ordered and (**B**) NiAs-type disordered (CrFe)S with a stoichiometric composition. The coherent potential approximation (CPA) method was adopted to calculate the disordered state. Partial DOSs refer to the angular-momentum-decomposed DOSs within the Voronoi cell surrounding each constituent atom. *SP*: spin polarization.

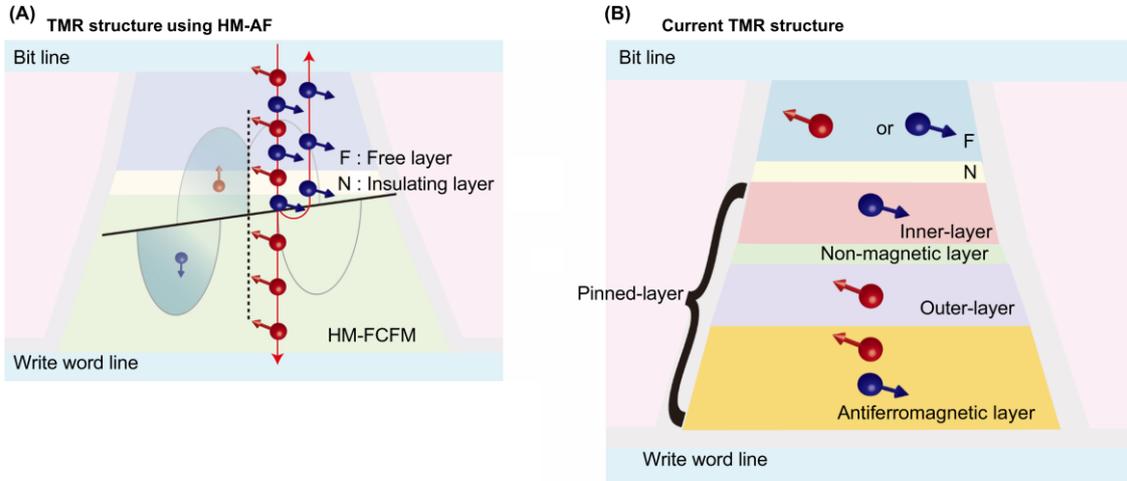

**Fig. 4. Illustration of a TMR device.** (**A**) A potential TMR device is constructed with a ferromagnetic free layer (F), non-magnetic (insulator) layer (N), and pinned layer of half-metallic fully compensated ferrimagnet (HM-FCFM). The HM-FCFM layer acts as an electron conductor in one spin orientation (up-spin, shown in red), but as an insulator in the opposite orientation (down-spin, shown in blue), as depicted in the schematic DOS for antiferromagnetic coupling. (**B**) Conventional tunneling magneto resistors (TRM) structure with a pinned layer comprising an inner layer, a non-magnetic layer (e.g., Ru-layer), an outer layer, and an antiferromagnetic layer. The device in (A), in which the HM-FCFM layer replaces the pinned layer in (B), facilitates a higher TMR ratio and decreased leakage current than the device in (B).

**Table 1. Physical properties of NiAs-type disordered (CrFe)S compound.**

| Structure (Space group) | Atom | Site Wyckoff position | Magnetic moment ($\mu_B$)[a] | Curie temperature (K)[a,b] | Ordering energy (eV)[a] | Spin polarization (%)[a] |
|---|---|---|---|---|---|---|
| CdI$_2$-type ordered (No. 156) | Fe | 1$a$ (0, 0, 0) | −2.94 (−2.86) | 1224 (1259) | – | 99.7 (99.7) |
| | Cr | 1$a$ (0, 0, 1/4) | 3.12 (3.03) | | | |
| | S | 1$b$ (1/3, 2/3, 1/4) 1$c$ (2/3, 1/3, 1/4) | −0.10 (−0.09) | | | |
| NiAs-type disordered (No. 194) | Fe | 2$a$ (0, 0, 0) | −3.06 (−2.98) | 1026 (1149) | 0.325 (0.356) | 99.7 (96.9) |
| | Cr | 2$a$ (0, 0, 0) | 3.27 (3.16) | | | |
| | S | 2$c$ (1/3, 2/3, 1/4) | 0.11 (−0.10) | | | |

[a] The magnetic moment, Curie temperature ($T_C$), ordering energy (i.e., the difference in the total energy between the CdI$_2$-type ordered structure and NiAs-type disordered structure), and spin polarization were calculated by the KKR method (see Supplementary Methods). The first values were obtained using the optimized lattice constants from first-principles calculations ($a$ = 0.3456 and $c$ = 0.5875 nm ($c/a$ = 1.700)), whereas the values in parentheses were obtained using the experimental lattice constants ($a$ = 0.3445 nm and $c$ = 0.5744 nm ($c/a$ = 1.667)).
[b] $T_C$ was calculated with mean field approximation using the exchange interaction, $J$, obtained in the paramagnetic state. The value of $J$ was calculated using Liechtenstein's method [37].

# Supplementary Materials for

## New type of half-metallic fully compensated ferrimagnet

S. Semboshi, R.Y. Umetsu*, Y. Kawahito, H. Akai*

**Materials and Methods**

Hexagonal NiAs-type (CrFe)S with an Fe/Cr atomic ratio of 1.0 is predicted to be a half-metallic fully compensated ferrimagnetic (HM-FCFM) material, primarily because it satisfies the condition of the total number of $d$-electrons per magnetic ion being 5 (i.e., the number of $d$-electrons in a pair of magnetic ions is 10). Based on the phase diagram of the Cr–Fe–S ternary alloy system presented in **Fig. S1A** [32, 39], a single phase of (CrFe)S has a wide compositional range at 1223 K with Cr (or Fe) content ranging from 0 to 50 at.% and S content ranging from 50 to 55 at.%. However, (CrFe)S with an Fe/Cr ratio of 1.0 is expected to decompose into two phases, i.e., Fe-rich (CrFe)S and Cr-rich (CrFe)S; three phases, i.e., Fe-rich (CrFe)S, Cr-rich (CrFe)S, and α-Fe (body-centered cubic (bcc)) at approximately 973 K (**Fig. S1B**); or three phases, i.e., Fe-rich (CrFe)S, $FeCr_2S_4$, and α-Fe (bcc) below 873 K (**Fig. S1C**).

In the preliminary experiment, we attempted to synthesize a stoichiometric $(Cr_{25}Fe_{25})S_{50}$ (at.%) specimen via a powder sintering procedure, as reported by Sokolovich and Bayukov [32]. However, the obtained sample had two phases, namely, Fe-rich (CrFe)S and $FeCr_2S_4$. Thus, we prepared a sample with a nominal composition of $(Cr_{23}Fe_{23})S_{54}$ to obtain a single-phase HM-FCFM material. This off-stochiometric (CrFe)S compound was synthesized by a powder-metallurgy process. Raw elemental powders of pure Fe (>99.9 wt.% purity, particle size <53 μm), Cr (99.9 wt.%, 180–300 μm), and S (99.99 wt.%, <75 μm) were weighed to achieve the nominal composition, and mixed well using a rotating mixer to obtain homogeneous mixed power. The mixed powder was packed into a die made of high-speed steel, and then compressed by cold uniaxial pressing under a pressure of >300 MPa at room temperature (293 K). In this procedure, since the Cr and Fe powder deformed plastically, we could obtain a high-density cylindrical compact with a diameter of 10 mm and height of 10 mm. The green compact was capsuled in a quartz tube under an Ar atmosphere using Ar gas of 99.99999% in purity with a pressure of 0.1 MPa and then sintered at 1253 K for 24 h to be fully homogenized. (According to the Cr–Fe–S phase diagram (**Fig. 1**), a sintering temperature of >1223 K is required.) Following this, it was quenched in water. The obtained sample was gently crushed into a rough powder using an agate mortar.

The appearance and compositional homogeneity of the sintered sample were examined by scanning electron microscopy (SEM; HITACHI S-3400N) with a conventional tungsten emitter and energy-dispersive X-ray spectroscopy (EDS) at an accelerating voltage of 15 kV. The composition of the sample was physically evaluated by EDS and chemically evaluated by inductively coupled plasma–atomic emission spectroscopy (ICP-AES) using a Thermo Fischer Scientific IRIS Advantage DUO instrument. The phase and structure of the sample were identified by powder X-ray diffraction (XRD) measurements (RIGAKU Rint-Ultima III) at room temperature with

Cu Kα radiation at 40 kV. The 2θ scan angle was 20–100° with a step size of 0.02°. The lattice parameters were determined by extrapolating the measured lattice parameters to θ = 0° as a function of $\cos^2\theta$. The phase stability of the sample was examined by differential scanning calorimetry (DSC) using SII EXSTAR 6000 from room temperature (293 K) to 1100 K at heating and cooling rates of 10 K/min under an 99.99999% Ar gas flow of 50 mL/min.

Magnetic measurements were carried out at a heating rate of 2 K/min using a superconducting quantum interference device (SQUID) magnetometer and a vibration sample magnetometer (VSM) equipped with a physical property measurement system (PPMS) produced by Quantum Design Ltd.

**Supplementary Text**
Crystal structure
   The powder XRD profile of the sample synthesized by sintering and quenching (**Fig. S2A**) showed that the sample had a hexagonal NiAs-type structure. To evaluate this result in detail, we simulated the XRD profiles based on possible NiAs-type (CrFe)S structures (**Figs. S2(B-E)**). **Figure S2B** shows the XRD profile corresponding to the stoichiometric $CdI_2$-type ordered structure illustrated in **Fig. 1C**, while **Fig. S2C** shows the stoichiometric NiAs-type disordered structure illustrated in **Fig. 1B**. There is no difference between the simulated XRD profiles of the $CdI_2$-type ordered structure and NiAs-type disordered structure. The experimental XRD profiles shown in **Fig. S2A** mostly fit the simulated profiles for both the stoichiometric $CdI_2$-type ordered and NiAs-type disordered structures. This suggests that it is difficult to evaluate the degree of ordering from the XRD analysis. However, Sokolovich and Bayukov [32] reported that similar but stoichiometric (CrFe)S specimens synthesized by sintering and quenching exhibited a NiAs-type disordered structure rather than a $CdI_2$-type ordered structure, as examined by Mössbauer spectra.

The chemical composition for the sample obtained by ICP-AES measurements was 22.7% Cr–23.3% Fe–54.0% S (at.%). This is nearly identical to the nominal composition of 23% Cr–23% Fe–54% S (at.%). Since the (CrFe)S sample synthesized in this study has a S-rich off-stoichiometric composition with an Fe/Cr ratio of approximately 1.0, the excess S atoms and/or vacancies ($Va$) are expected to be distributed in the 2$a$ sites in order to retain a single-phase NiAs-type structure. **Figs. S2D** and **E** present off-stoichiometric NiAs-type structures with the composition $(Cr_{23}Fe_{23})S_{54}$. **Fig. S2D** is obtained by assuming that the excess S atoms (4 at.%) are randomly substituted in the 2$a$ sites (i.e., the sample can be represented as $(Cr_{23}Fe_{23}S_4)S_{50}$), while **Fig. S2E** is determined by assuming that 7.4 at.% of vacancies ($Va$) are substituted in the 2$a$ sites (i.e., $(Cr_{21.3}Fe_{21.3}Va_{7.4})S_{50}$, in which the ratio of Cr, Fe, and S atoms is 23:23:54). The experimental diffraction intensities for 010 and 011 in **Fig. S2A** agree well with the simulated patterns for the $(Cr_{21.3}Fe_{21.3}Va_{7.4})S_{50}$ lattice shown in **Fig. S2E**, rather than the simulated patterns for the $(Cr_{23}Fe_{23}S_4)S_{50}$ lattice shown in **Fig. S2D**. This indicates that the S-rich (CrFe)S compound synthesized in this study potentially has a NiAs-type disordered structure containing a certain number of vacancies in the 2$a$ sites. It may be noted that 7.4% of vacancies in the 2$a$ sites might be excessive for retaining the crystal structure and electron state of the NiAs-type lattice. Therefore, it may be reasonably assumed that the 2$a$ sites in the experimental samples

are occupied not only by the vacancies, but also by a portion of excess S atoms along with Cr and Fe atoms. To elucidate the detailed structure and degree of ordering of the (CrFe)S compound synthesized, neutron diffraction analysis is ultimately required, which will be reported in the future.

Phase stability

The thermal stability of the (CrFe)S sample was examined by DSC measurements, and the results are presented in **Fig. S3**. The DSC curves observed during the first and second heating and cooling cycles at sweep rates of 10 K/min are indicated. During the first heating process, a relatively strong exothermic peak was observed at approximately 700 K (marked by a solid circle in **Fig. S3**). This peak likely corresponds to the decomposition of the metastable (CrFe)S phase obtained by quenching into three stable phases, namely, Fe-rich (CrFe)S, α-Fe, and $FeCr_2S_4$, as expected from the phase diagram in **Fig. S1C** [39]. We also observed reversible weak endothermic and exothermic peaks during heating and cooling, respectively, beginning at 910 K during the first and second DSC curves. The synthesized $Cr_{23}Fe_{23}S_{54}$ compound exhibits three phases below 873 K (**Fig. S1C**), but only a single phase of (CrFe)S above 973 K (**Fig. S1B**). Thus, the three phases of Fe-lean (CrFe)S, α-Fe, and $FeCr_2S_4$ that are formed by decomposition at approximately 700 K are expected to transform back to (CrFe)S at 910 K, as an equilibrium reaction. In addition, the phase transformation occurs in the equilibrium state, which causes the reversible reaction of the weak endothermic and exothermic peaks around 910 K after the second heating and cooling processes. The DSC curves suggest that quenching from a temperature higher than 910 K would be required to obtain a single-phase (CrFe)S compound with an Fe/Cr ratio of 1.0. In addition, a small step indicated by a double arrow is observed at approximately 500 K only during the first heating process. This is associated with the Curie temperature of the (CrFe)S sample and corresponds to the thermomagnetization curve ($M$–$T$ curve) (see **Fig. 2A**). Since the phase is decomposed during the first heating process, no step is observed during the second heating process.

Theoretical calculation of the density of states

First-principles calculations were performed using the density functional theory framework, within the local density approximation/generalized gradient approximation (LDA/GGA), using the all-electron full-potential Korringa–Kohn–Rostoker Green's function method (FPKKR) [40] combined with the coherent potential approximation (CPA) [41, 42]. The CPA procedure is needed to simulate realistic situations where the S atoms occupy some of the transition metal sites ($2a$ sites), as a result of which, the system is not stoichiometric anymore. We employed the GGA scheme proposed by Perdew, Burke, and Ernzerhof (PBE) [43], although the results did not depend much on the choice of the LDA/GGA schemes. We assumed a $CdI_2$-type ordered structure for the systems where a Cr atom occupies one of the two transition metal sites in the NiAs-type structure, while an Fe atom occupies the other transition metal site in each unit cell, as observed in **Fig. 1C**. We also considered the situation where the Cr and Fe atoms occupy two transition metal sites randomly with a probability of occupation of 0.5 (this is the case for the NiAs-type structure presented in **Fig. 1B**). In such cases, we applied the FPKKR-CPA to take the random arrangements into account. The same procedure was employed when the partial substitutions of the Cr and Fe atoms for S atoms and vacancies were considered.

The Curie temperatures of the systems were calculated by constructing a low energy effective Hamiltonian using Liechtenstein's formula [37]. Once the effective Hamiltonian is obtained, which takes the form of the Heisenberg model with fairly long-range coupling constants $J_{ij}$s, we can estimate the Curie temperature within the mean field approximation. Although the mean field approximation generally overestimates the Curie temperature, the drawback is not particularly impactful for the present system owing to the highly concentrated magnetic ions.

It is important for first-principles calculations to use proper crystallographic parameters such as lattice parameters and atomic arrangement of the unit lattice. The lattice parameters of stoichiometric NiAs-type (CrFe)S are predicted to be $a = 0.3456$ nm and $c = 0.5875$ nm based on first-principles calculations. The DOS is also drawn, as shown in **Fig. 3**. The lattice parameters of the (CrFe)S sample in this study, however, were measured to be $a = 0.3446$ nm and $c = 0.5743$ nm based on the XRD profile shown in **Fig. S2A**, with an expected experimental accuracy of ±0.0010 nm. When we performed the first-principles calculations using the experimental lattice parameters, we obtained the DOS for the stoichiometric $CdI_2$-type ordered and NiAs-type disordered (CrFe)S compounds, as shown in **Fig. S4,** as well as the values of physical properties such as magnetic moment and Curie temperature, as listed in **Table 1**. We find that these are not significantly different from the theoretical values obtained by first-principles calculations, presented in **Fig. 3 and Table 1**.

Since the (CrFe)S sample synthesized in this study has an off-stoichiometric composition, the atomic arrangement cannot be strictly determined from the XRD profile, as discussed above. We assume that the 2*a* sites in the NiAs-type structure are randomly occupied by Cr, Fe, excessive S atoms, and vacancies (*Va*). For example, arbitrarily assuming that the structure is represented as $(Cr_{21.8}Fe_{21.8}S_{1.4}Va_{5.0})S_{50}$, the DOS is calculated as shown in **Fig. S5**. We can see that the DOS shown in **Fig. 5** is also essentially the same as the DOS shown in **Figs. 4** and **S4**. Thus, it is concluded that the half-metal-type electronic state is retained, as long as the total number of *d*-electrons per magnetic ion is 5 (when the Fe/Cr ratio of the (CrFe)S compound is 1.0 in the case of the NiAs-type (CrFe)S compound), even if the lattice parameters fluctuate slightly and the S content deviates from the stoichiometric composition. Based on this conclusion, it is expected that (CrFe)S compounds with S content ranging from 50 to 55 at.%, including the $Cr_{23}Fe_{23}S_{54}$ compound synthesized in this study, might exhibit half-metallic fully compensated ferrimagnetism, which will be confirmed in the future.

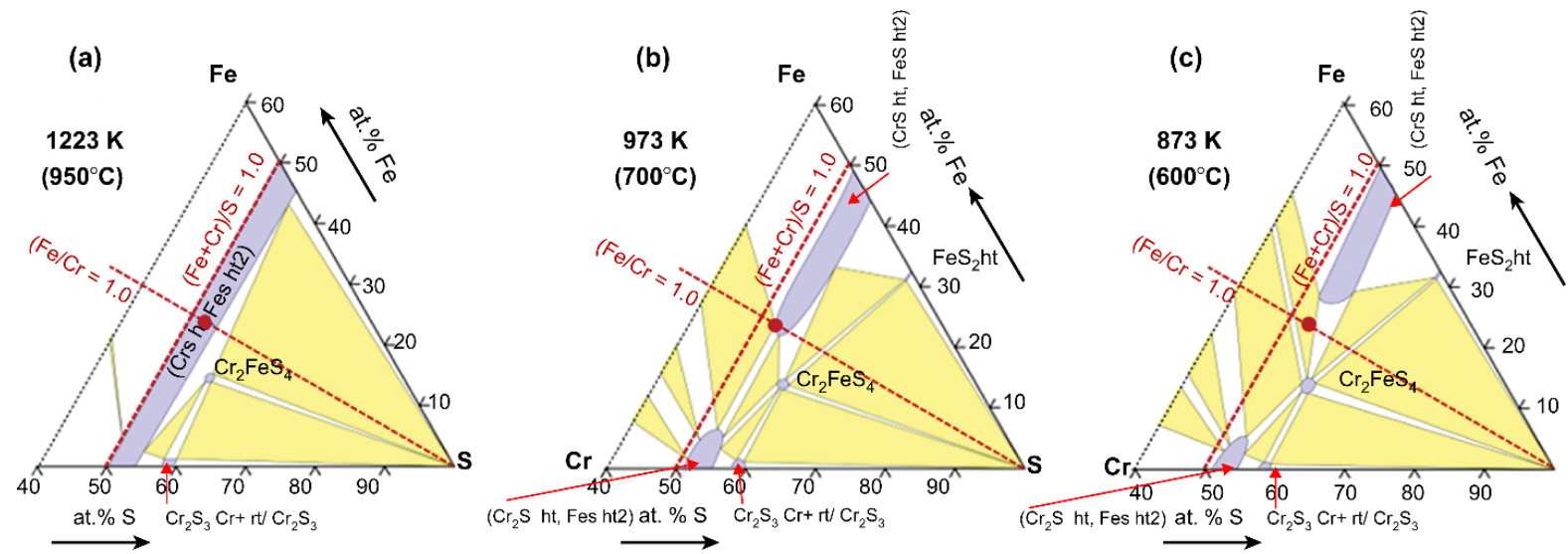

**Fig. S1.**
Partial phase diagram of the Cr–Fe–S ternary system at (**A**) 1223 K, (**B**) 973 K, and (**C**) 873 K [39]. (CrFe)S with the composition $(Cr_{23}Fe_{23})S_{54}$ (marked by the red dot) was synthesized by sintering and quenching in this study.

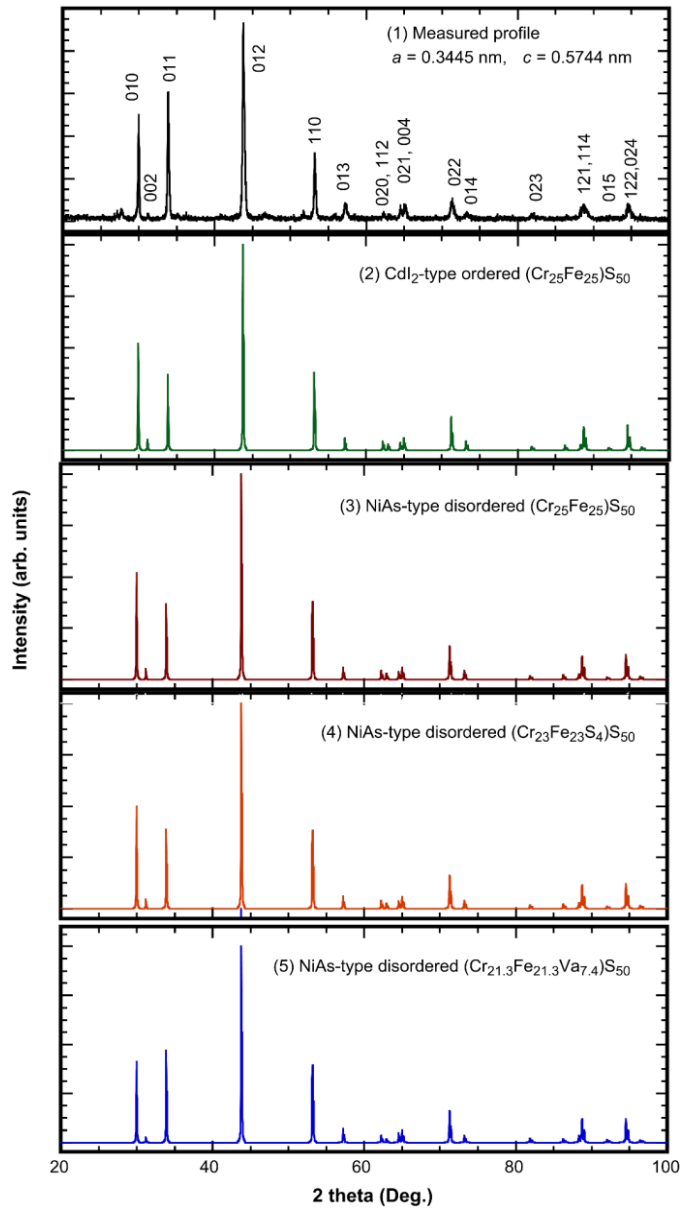

**Fig. S2.**
(**A**) Experimental X-ray diffraction profile of specimen synthesized in this study, and (**B**–**E**) simulated profiles for (**B**) CdI$_2$-type ordered structure with stoichiometric composition, with Cr (25 at.%) and Fe (25 at.%) ordered in the 2$a$ (0, 0, 0) sites and S (50 at.%) in the 2$c$ sites for the Wyckoff position (shown in **Fig. 1C**); (**C**) NiAs-type disordered structure with stoichiometric composition (**Fig. 1B**); (**D**) NiAs-type disordered structure with an off-stoichiometric (Cr$_{23}$Fe$_{23}$S$_4$)S$_{50}$ (at.%) composition, assuming that Cr, Fe, and excess S atoms are randomly distributed in the 2$a$ sites; and (**E**) NiAs-type disordered structure with an off-stoichiometric (Cr$_{21.3}$Fe$_{21.3}$$Va_{7.4}$)S$_{50}$ composition, assuming that Cr atoms, Fe atoms and 7.4 at.% of vacancies ($Va$) are randomly distributed in the 2$a$ sites.

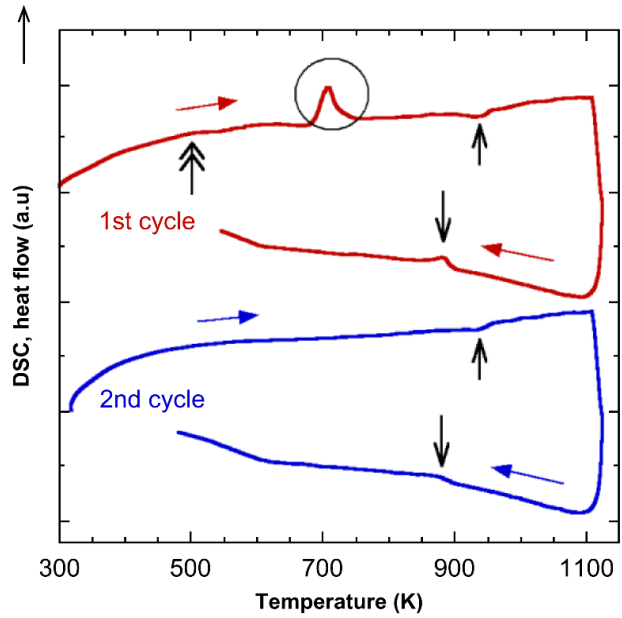

**Fig. S3.**
Cyclic DSC heating and cooling curves for (CrFe)S synthesized by sintering and quenching, measured at the heating and cooling rates of 10 K/min. The exothermic peak at 700 K in the first heating curve (marked by solid circle) was caused by decomposition of NiAs-type (CrFe)S phase. The single arrows at approximately 910 K and double arrow at 500 K correspond to the re-composition of α-Fe and $FeCr_2S_4$, and the Curie temperature, respectively.

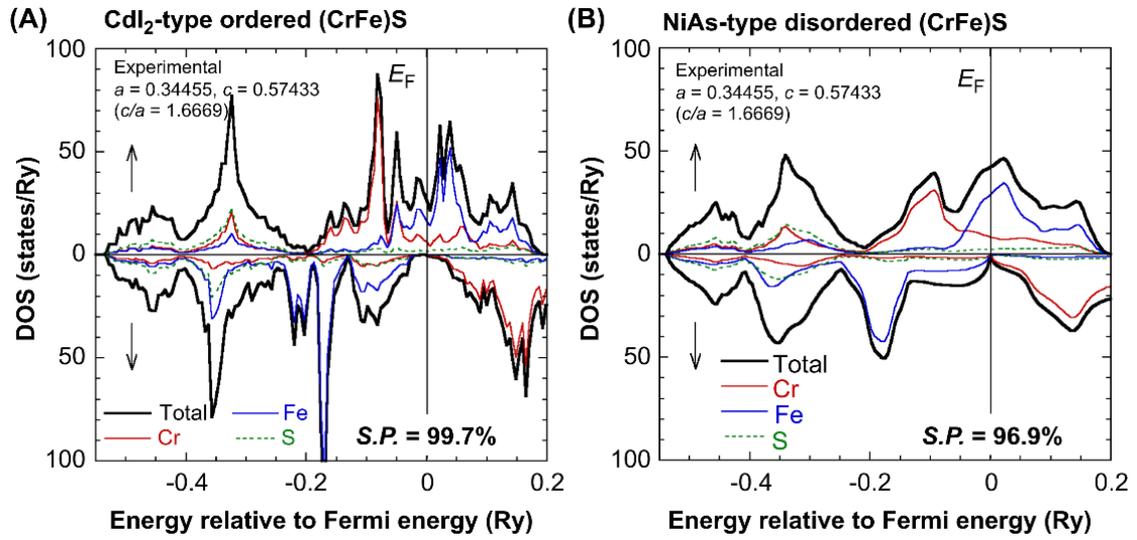

**Fig. S4.**
Partial DOS of Cr, Fe, and S atoms and total DOS of (**A**) CdI$_2$-type ordered and (**B**) NiAs-type disordered (CrFe)S with stoichiometric composition, calculated using the lattice parameters obtained from the experimental XRD pattern shown in **Fig. S2A**. The ordered and disordered states calculated employed coherent potential approximation (CPA).

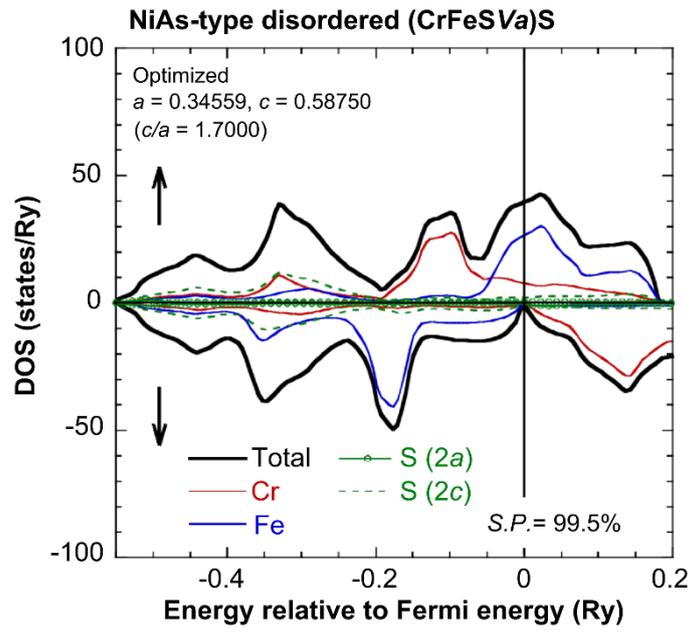

**Fig. S5.**
Partial DOS of Cr, Fe, and S and the total DOS of an off-stoichiometric NiAs-type disordered $(Cr_{21.8}Fe_{21.8}S_{1.4}Va_{5.0})S_{50}$.